\documentstyle[amssymb,aps,prb,twocolumn]{revtex}

\newcommand{\be}{\begin{eqnarray}}
\newcommand{\ee}{\end{eqnarray}}

\begin{document}

\twocolumn[\hsize\textwidth\columnwidth\hsize\csname @twocolumnfalse\endcsname

\title{Thermoelectric power of MgB$_{2-x}$Be$_x$}
\author{J. S. Ahn$^{1}$, E. S. Choi$^2$, W. Kang$^2$, D. J. Singh$^3$, and E. J. Choi$^{1,4}$}
\address{$^1$Center for Strongly Correlated Materials Research, Seoul National\\
University, Seoul 151-742, Republic of Korea}
\address{$^2$Department of Physics, Ewha Womans University, Seoul 120-750, Republic\\
of Korea}
\address{$^3$Center for Computational Materials Science, Naval Research Laboratory,\\
Washington, DC 30375, U.S.A.}
\address{$^4$Department of Physics, University of Seoul, Seoul 130-743, Republic of\\
Korea}
\maketitle

\begin{abstract}
We investigated thermoelectric power $S(T)$ of MgB$_{2-x}$Be$_{x}$ ($x=0$,
0.2, 0.3, 0.4, and 0.6). $S(T)$ decreases systematically with $x$,
suggesting that the hole density increases. Our band calculation shows that
the increase occurs in the $\sigma $-band. With the hole-doping, $T_{c}$
decreases. Implication of this phenomenon is discussed within the BCS
framework. While the Mott formula explains only the linear part of $S(T)$ at
low temperature, incorporation of electron-phonon interaction enables us to
explain $S(T)$ over wide temperature range including the anomalous behavior
at high temperature.
\end{abstract}

\pacs{PACS number: }
\date{\today}

\vskip2pc]


\section{Introduction}

Since the recent discovery of superconductivity in MgB$_{2}$,\cite
{Akimitsu2001} both theoretical and experimental efforts have been made to
understand its structural and electrical properties. Most theoretical works
suggest that coupling of the $\sigma $-hole with B-plane phonon is the key
ingredient of the superconductivity.\cite{Kortus2001,An2001,Liu2001} Boron
isotope effect\cite{Budko2001} and other experimental data showed that the
material is in the intermediate or strong BCS coupling regiem ($\lambda
=0.7\sim 1$).\cite
{Slusky2001,Bouquet2001,Osborn2001,Lorenz2001a,Goncharov2001,Kang2001}

Chemical substitution experiment has drawn much attention due to the
possibility of obtaining higher superconducting transition temperature $%
T_{c}.$ In general, by replacing Mg- or B-sites with other elements,
physical parameters such as lattice constants and carrier density change.
Therefore, study of $T_{c}$ shift in well-controlled substitution samples
provides a chance to understand the superconductivity in detail. For the
Mg-site substitution, several compounds such as Mg$_{1-x}$Al$_{x}$B$_{2}$
have been studied.\cite{Slusky2001,Zhao2001,Kazakov2001,Moritomo2001} For
the B-site, C- and Be-substitutions have been reported.\cite{Ahn2001,AhnBe}

It is important to know how these substitutions change the carrier density
and how the change is related with $T_{c}$ shift. Thermoelectric power (TEP)
measurement is a useful probe of the carrier density. For example, TEP
experiment on Mg$_{1-x}$Al$_{x}$B$_{2}$ showed that the Al-substitution
dopes electrons.\cite{Lorenz2001} Theoretical calculation on Mg$_{1-x}$Al$%
_{x}$B$_{2}$ suggested that the doping occurs largely in the $\sigma $-hole
pocket.\cite{Suzuki2001} In the present paper, we report our results of TEP
measurement and theoretical band calculation on MgB$_{2-x}$Be$_{x}$ samples
where B is substituted with Be. In an earlier paper, we showed that in this
compound, the MgB$_{2}$ phase is maintained up to $x=0.6.$ Also the lattice
constants and $T_{c}$ changed systematically with Be-substitution.\cite
{AhnBe} We find that the TEP decreases with $x$, opposite to that in Mg$%
_{1-x}$Al$_{x}$B$_{2}$, which suggests that hole is doped. Also, changes in
the band structure indicate that the hole doping occurs mostly in the $%
\sigma $-band. Thus, Be-substitution brings about hole-doping into the boron
plane. We consider the consequence of this effect on the transition
temperature and arue that the hole density change plays only a second role
in the $T_{c}$ shift.

Meanwhile, it is well known now that the temperature dependence of TEP is
one of the unconventional features of MgB$_{2}$: At low $T,$ TEP is linear
in $T$ which is normal for most metals, but crosses over to a sublinear
behavior as $T$ increases. There have been many interpretations of this
phenomenon. \cite{Lorenz2001,EChoi2001,Putti2001} We find that the crossover
exhibits a systematic change with the Be-substitution. Furthur, we show that
TEP in the whole $T$ region is explained by a single model function in which
the electron-phonon interaction is explicitely taken into account.

\section{Experimental}

Polycrystalline samples used in this experiment were synthesized by a powder
metallurgical technique using a high pressure furnace.\ Starting materials
are fine powders (-325 mesh) of Mg (99.8\%, Alfa Aesar), amorphous B
(99.99\%, Alfa Aesar), and Be (99.+\%, Alfa Aesar). Stoichiometric amounts
of powders are mixed and pelletized. The pellets are placed in a tungsten
vessel with a close-fitting cap, reacted two hour at 850 $^{\circ }$C under
20 atm. of high purity argon atmosphere.

In MgB$_{2-x}$Be$_{x},$ the MgB$_{2}$ phase is maintained up to $x=0.6$.
Their structural and superconducting properties were reported elsewhere.\cite
{AhnBe} We found that as a function of Be-substitution, lattice parameters
show decreasing $a$- and increasing $c$-values, and transition temperature $%
T_{c}$ decreases as summarized in Table I.

For TEP measurements, bar-shaped samples (with dimensions of $\sim $ 4$%
\times $0.5$\times $0.1 $mm^{3}$) were mounted on two resistive heaters.
Gold wires were used for thermoelectric potential leads. Chromel-constantan
thermocouples were used for the temperature gradient measurement. Sample
ends and thermocouple beads were glued to the heater blocks by Stycast
epoxy. In our measurement, thermopower from the contact wires was carefully
calibrated.

\section{Result and Discussion}

Figure 1 shows thermoelectric power $S(T)$ of MgB$_{2-x}$Be$_{x}$ samples.
For MgB$_{2}$ ($x=0$), $S(T)$ is linear in $T$ at low temperature above $%
T_{c}$. At higher $T$ ($>150$ K), it crosses over to a sub-linear behavior.
These features together with the large jump at $T_{c}$ ($\sim $ 1.4 $\mu $%
V/K) are consistent with the earlier reports.\cite{Lorenz2001,EChoi2001} As
boron is substituted with Be, the linear slope decreases. Also the crossover
temperature is lowered. At $x=0.6$, $S(T)$ changes sign at a low temperature.

Thermal diffusion of carriers gives rise to the linear-in-$T$ behavior in $%
S(T)$ at low temperatures. Kinetic transport theory shows that $S(T)=(\pi
^{2}k_{B}^{2}T/3e)\sigma ^{^{\prime }}(E)/\sigma (E)$, where the
dc-conductivity $\sigma (E)$ and its derivative $\sigma ^{^{\prime }}(E)$
are calculated at the Fermi energy. \cite{Blatt1976} In a single parabolic
band system, it is approximated to the Mott formula:\cite{Blatt1976,Mott1993}
\begin{equation}
S(T)=X_{b}T=\pm \left( \frac{\pi ^{2}k_{B}^{2}}{3eE_{F}}\right) T,
\end{equation}
where $k_{B}$ is the Boltzman constant, $\pm \ $corresponds to the carrier
sign, and $E_{F}$ is the Fermi energy relative to the band maximum (or
minimum). However, MgB$_{2}$ is a multi band system where $\sigma $- and $%
\pi $-bands coexist. Also the Fermi surface is not spherical. Thus, Eq. (1)
can be applied only approximately here. The positive value of $X_{b}$ in
pristine MgB$_{2}$ suggests that the dominant carrier is hole. The decrease
of $X_{b}$ with $x$ suggests that the hole density increases. To obtain $%
X_{b}$, we fit the linear part of $S(T)$ with Eq. (1) as shown by the dashed
lines. Since the data are not extrapolated to zero, vertical shifts were
needed in the fit.\cite{mention}

To understand the behavior of $X_{b}$ more quantitatively, we calculated
band structures of MgB$_{2-x}$Be$_{x}$ as shown in Fig. 2. Here we used the
local density approximation (LDA) with the linearized augmented plane wave
method. To account for the random substitution of Be, the virtual crystal
approximation (VCA) was employed. Details of the calculation method were
described by Mehl et al.\cite{Mehl2001} In current work, the experimental
lattice parameters in Table I were used. Four bands contribute to the Fermi
surfaces: two $\sigma $-bands with B $p_{x,y}$ character give the 2D
hole-type cylinders around the $\Gamma $-$A$ line and two $\pi $-bands form
the 3D honeycomb tubular networks. The latter consists of one electron-type
at the $H$-point and another hole-type at the $K$-point. Note that the most
prominent change with $x$ is the hole increase in the $\sigma $-bands.
Change in the $\pi $-bands is relatively small. These results tell us that
the Be-substitution dopes hole into the $\sigma $-bands.

It is interesting to note that $T_{c}$ decreases with the hole-doping,
similarly to the electron-doped Mg$_{1-x}$Al$_{x}$B$_{2}$. Theoretical
studies show that the superconductivity arise from the $\sigma $-band holes
coupled with the B-plane phonons. $T_{c}=39$ K in MgB$_{2}$ can be produced
from the McMillan formular using the coupling constant $\lambda =1.01$ and
the Coulomb pseudopotential $\mu ^{\ast }=0.13$.\cite{An2001,Liu2001} Here $%
\lambda $ is proportional to the density of state (DOS) of the $\sigma $%
-bands at the Fermi energy, $N_{h}(E_{F}).$ From our band calculation, we
find that $N_{h}(E_{F})$ =0.22 (eV$^{-1}$ per cell) at $x=0$ and $%
N_{h}(E_{F})$ =0.24 (eV$^{-1}$ per cell) at $x=0.6.$ (the $\sigma $-bands
are highly 2D-like and the DOS increases only slightly with hole doping.)
Then $\lambda $ will increase proportionally to become 1.09 at $x=0.6,$ if
we assume the other parameters determining $\lambda $ do not change. (see
for example Eq. (3) of Ref. 3) This yields $T_{c}=45$ K, which is in sharp
contrast with the observed $T_{c}$ decrease. This suggests that the other
parameters change significantly with the substitution and their effects
overcome the $N_{h}(E_{F})$ contribution. In another paper,\cite{AhnBe} we
dealt with this issue and showed that the lattice constant change is the
primary cause of the reduced superconductivity.

Now we consider the unusual behavior of $S(T)$, i.e., the deviation from the
linear dependence at high $T$. In previous works, it has been attributed to
the minor carrier contribution,\cite{Lorenz2001} the thermally activated
transport,\cite{EChoi2001} and to the phonon-drag effect.\cite{Putti2001}
Here, we consider effect of electron-phonon interaction on $S(T).$ According
to Kaiser, the interaction contributes to enhance the TEP through modifying
the carrier mass and thus the thermal diffusion.\cite{Kaiser1984} Taking
this effect into account, Eq. (1) is rewritten as 
\begin{equation}
S(T)=[1+\lambda \bar{\lambda}_{s}(T)]X_{b}T,
\end{equation}
where $X_{b}$ is the slope in Eq. (1), $\lambda $ is the electron-phonon
coupling constant, and $\bar{\lambda}_{s}(T)$ is a function which represents
the $T$-dependent thermopower enhancement: 
\begin{equation}
\bar{\lambda}_{s}(T)=\int_{0}^{\infty }d\omega \frac{\alpha ^{2}F(\omega )}{%
\omega }G_{s}(\frac{\hbar \omega }{k_{B}T}).
\end{equation}
Here, the normalized Eliashberg function $\alpha ^{2}F(\omega )$ consists of
the phonon density of states $F(\omega )$ and the coupling constant $\alpha $%
. $G_{s}(\hbar \omega /k_{B}T)$ is a function associated with thermal
population of phonons. For MgB$_{2},$we calculated $\bar{\lambda}_{s}(T)$
using $\alpha ^{2}F(\omega )$ reported by Liu et al.\cite{Liu2001,Debye} and
fit the data with Eq. (2). Here we used $\lambda $ and $X_{b}$ as fitting
parameters. For $x>0,$ the Be-substitution into the B-plane will change,
probably significantly, the phonon structure. Thus $\alpha $ and $F(\omega )$
will depend on $x$. Since they are not known$,$ we took the values of MgB$%
_{2}$ in calculating $\bar{\lambda}_{s}(T)$. Thus $\lambda $ and $X_{b}$ we
estimate for $x>0$ samples are under large uncertainties.

Figure 3(a) shows the fit for MgB$_{2}$ (solid line). The bare diffusion
part ($X_{b}T$) and the enhancement part ($\lambda \bar{\lambda}%
_{s}(T)\times X_{b}T$) are represented with dashed and dash-dotted line,
respectively. The latter has a broad maximum at $T\sim 215$ K. Inset shows
calculated behavior at higher temperature. Fig. 3(b) shows the fitting
results for MgB$_{2-x}$Be$_{x}$. Note that the fit is reasonable except the
small deviation for $x=0.3$.

In Fig. 4, we summarize the linear slope $X_{b}$ obtained from our analyses.
The fitting results using Eq. (1) and Eq. (2) are shown with the
filled-circles and the triangles, respectively. Note that $X_{b}$ from the
modified model is smaller than that from the bare diffusion model. This is
due to the enhancement effect contained in Eq. (2). Also, $\ $we estimated $%
X_{b}$ from the band calculation (the dash-dotted line). Here the $E_{F}$ in
Eq. (1) was taken from the $\sigma $-hole bands, assuming \ contributions
from the other bands are negligible.\cite{piband} The band calculation
result is closer to the modified model result at low doping region, $x\leqq
0.3$, which supports the importance of the electron-phonon interaction
effect. At $x=0.3$, $X_{b}$ from the two fits exhibit a sudden drop. This
drop may be related to the observed structural change in the same
composition.\cite{AhnBe} The incomplete agreement between the fit result and
the band calculation result may come from complex effects not included in
this work such as the multi-band contributions and anisotropic transport.

Regarding the electron-phonon coupling constant (EPC), we obtain $\lambda
=0.90$ for\ MgB$_{2}$. This is in good agreement with the earlier reports of
0.7 $\sim $ 1.0.\cite{Kortus2001,An2001,Liu2001,Osborn2001} For $x>0$, $%
\lambda $ increases to 0.98($x=0.2$), 1.31(0.3), 1.34(0.4), 1.47(0.6). This
result is quite unusual because, as $T_{c}$ decreases with $x,$ $\lambda $
is expected to decrease. Recently, evidences show that MgB$_{2}$ has two
gaps. In this case, EPC from transport measurement (= $\lambda _{tr})$ is
different from the EPC which determines the superconducting $T_{c}$ (= $%
\lambda _{sc}$).\cite{Liu2001} Thus, the increase of TEP $\lambda $ (= $%
\lambda _{tr}$) does not necessarily contradict with the $T_{c}$ decrease.
One should also keep in mind that the increase of TEP $\lambda $ may be
simply an errorneous effect that arise from the uncertainties in $\alpha $
and $F(\omega )$ for $x>0$ we mentioned above.\cite{extra}

Now, let us consider the sign change in $S(T)$ of $x=0.6$. Sign change in
TEP is widely observed in many alloyed systems, for example, Ag-Au alloy, 
\cite{Blatt1976} YBa$_{2}$Cu$_{3}$O$_{7-\delta }$,\cite{Lee1988} NbN$_{x}$, 
\cite{Siebold1993}, etc. In YBa$_{2}$Cu$_{3}$O$_{7-\delta }$, the change is
observed as oxygen deficiency $\delta $ increases. In NbN$_{x}$, $S(T)$ is
composed of the diffusive (positive in sign) and phonon-drag terms (negative
in sign). At high $T$, the former is dominant while the latter prevails at
low $T$. In the intermediate $T$ , sign change occurs.\cite{Blatt1976} It is
tempting to interpret our observation similarly: the negative $S(T)$ may
correspond to the phonon-drag effect. However, note that in the pristine MgB$%
_{2}$, the phonon-drag feature is not observed. Furthur, the feature, if
any, should be suppressed with Be-substitution because the phonon-drag
generally disappears as randomness is increased. Origin of the sign change
is thus remains for future study.

\section{Conclusion}

>From the TEP measurement and band structure calculation on MgB$_{2-x}$Be$%
_{x} $\ ($x=0$, 0.2, 0.3, 0.4, and 0.6), we found that the hole density
increases with $x$ in the $\sigma $-bands . Thus, the Be-substitution dopes
hole into the boron plane. The fact that $T_{c}$ shifts in the same
direction (lowering) as the electron-doped case suggests that carrier doping
is not the primary route to control the transition temperature in MgB$_{2}.$
This result is consistent with the 2D nature of the $\sigma $-bands.
Furthur, we showed that the anomalous behavior of TEP\ at high temperature
can be explained.by taking the electron-phonon interaction effect into
account.

\section*{Acknowledgments}

This work was supported by KRF-99-041-D00185 and by the KOSEF through the
CSCMR.

\begin{table}[h]
\caption{Physical properties of MgB$_{2-x}$Be$_x$. 
Transition width $\Delta T_c$ is determined from 10-90\% transition.\cite{AhnBe} }
\begin{tabular}{ccccc}
$x$ & $a$ ({\rm \AA})& $c$ ({\rm \AA})& $T_c$ (K) & $\Delta T_c$ (K) \\ \hline
0 & 3.084 & 3.522 & 38.4 & 1.2 \\
0.2 & 3.078 & 3.540 & 36.0 & 2.5 \\
0.3 & 3.073 & 3.597 & 33.0 & 5.5 \\
0.4 & 3.073 & 3.632 & 21.0 & 4.0 \\
0.6 & 3.062 & 3.639 & 8.4 & 1.5
\end{tabular}
\end{table}
%

\begin{figure}[tbph]
\caption{Themoelectric power $S(T)$ of MgB$_{2-x}$Be$_{x}$ ($x=0$, 0.2, 0.3,
0.4, and 0.6). Dashed lines represent linear fits to the data.}
\label{Fig:1}
\end{figure}

\begin{figure}[tbph]
\caption{LDA virtual crystal band structures of MgB$_{2-x}$Be$_{x}$ for $x=0$
(top), 0.3 (middle), and 0.6 (bottom). The experimental lattice parameters
are used. The horizontal reference at 0 denotes $E_{F}$.}
\label{Fig:2}
\end{figure}
\begin{figure}[tbph]
\caption{(a) Thermopower data of MgB$_{2}$ (open circles) and fit with the
modified diffusion model (solid line). Dashed- and dash-dotted lines
represent the diffusion and enhancement part, respectively. Inset shows the
calculated behavior at higher temperature. (b) TEP data of MgB$_{2-x}$Be$%
_{x} $(open circles) and the modified diffusion fit (solid lines).}
\label{Fig:3}
\end{figure}

\begin{figure}[tbp]
\caption{The linear slope $X_{b}$ of MgB$_{2-x}$Be$_{x}$. $\bullet $ :
determined from the Eq. (1). $\blacktriangle $ : from the modified diffusive
fit in Eq. (2). Theoretical results from the band calculations are shown
with dash-dotted line. Solid lines are for eye-guide. }
\label{Fig:4}
\end{figure}


\begin{references}
\bibitem{mention}  The non-zero crossing of the extrapolated TEP does not
agree with Eq.(1) (and also Eq.(2) below). This suggests that additional
contribution(s) (such as phonon-drag effect) to TEP may exist at low
temperatures.
\bibitem{piband}  The $\pi $-hole and the $\pi $-electron bands have similar
3D structure. Thus, their contributions to $S(T)$ may partly cancel each
other.
\end{references}
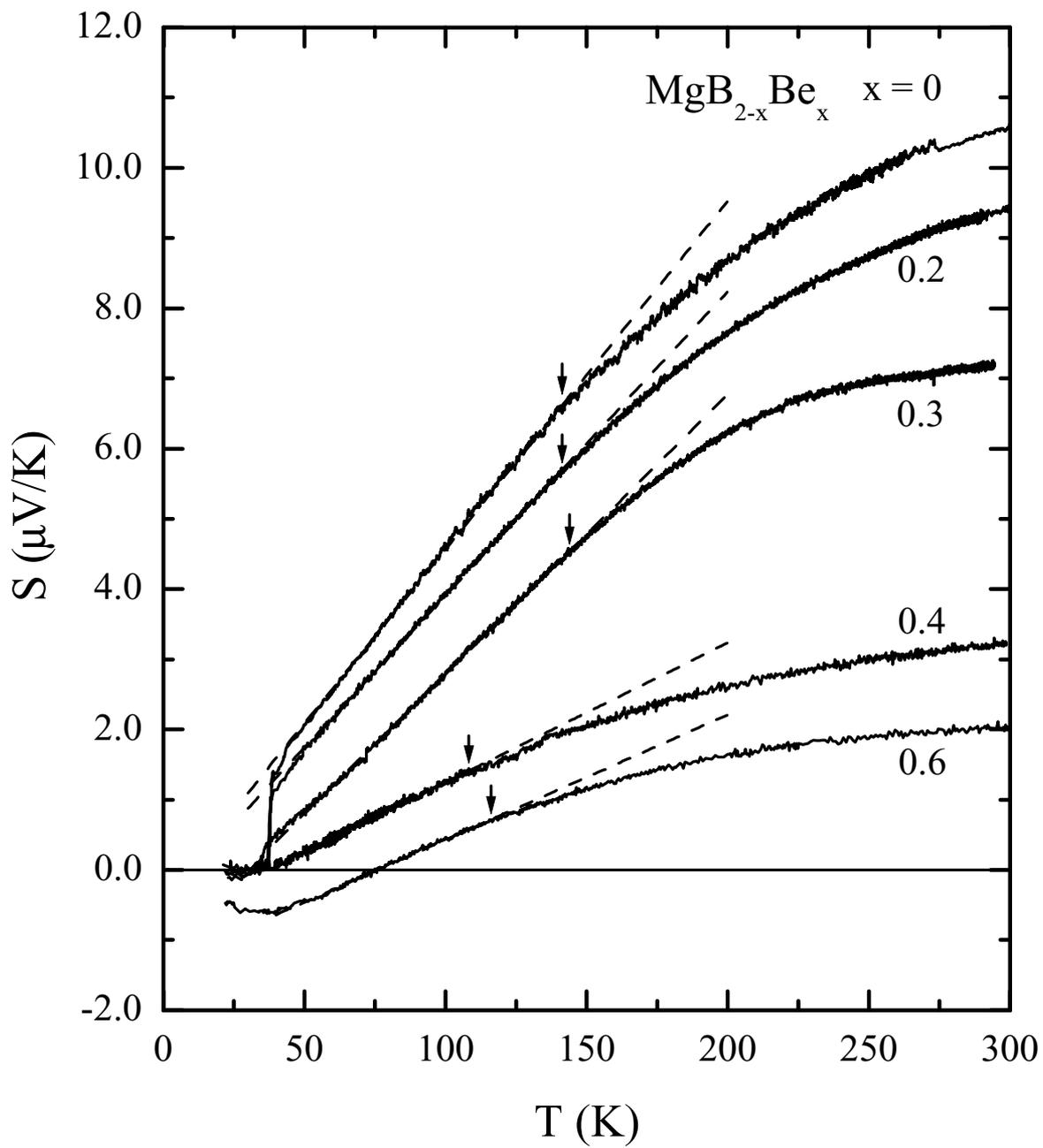
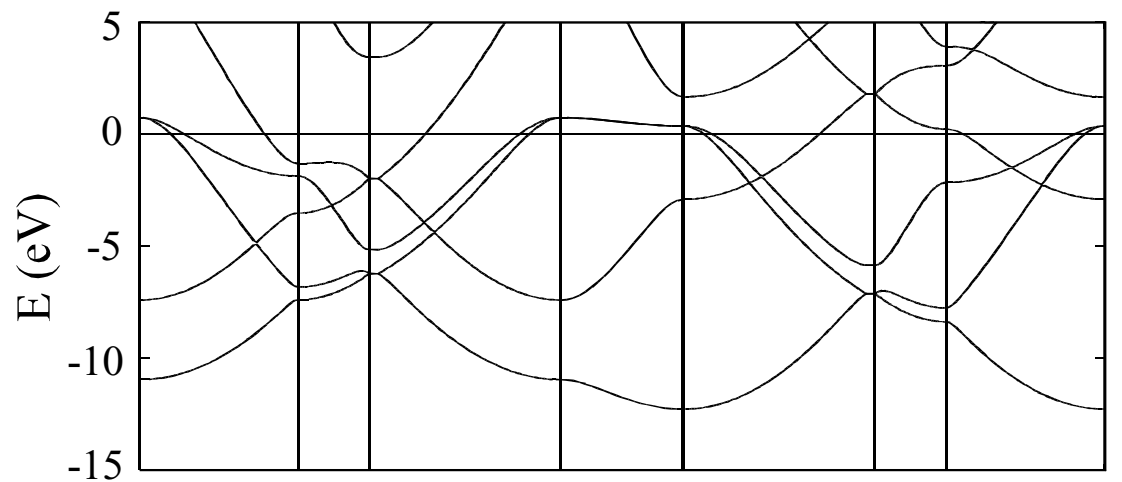
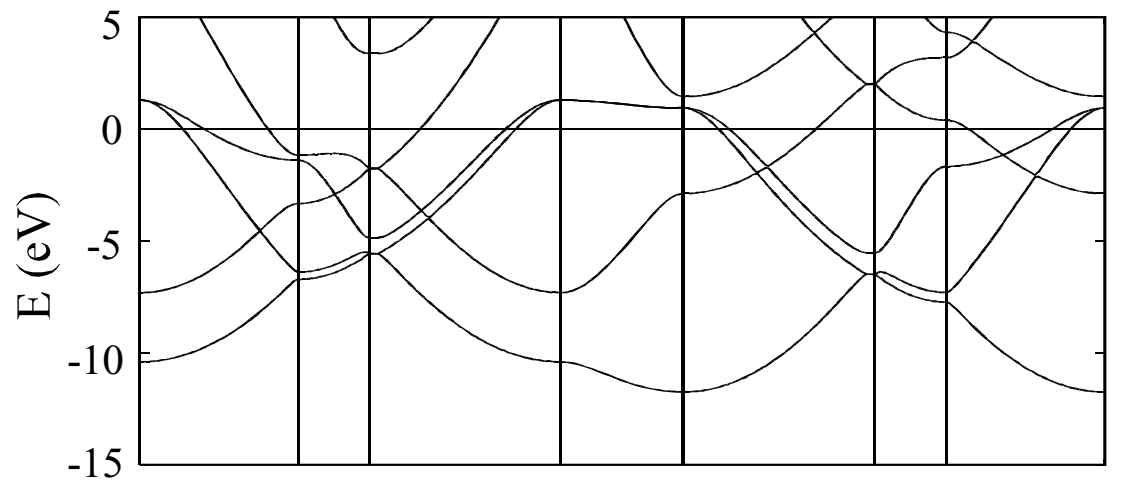
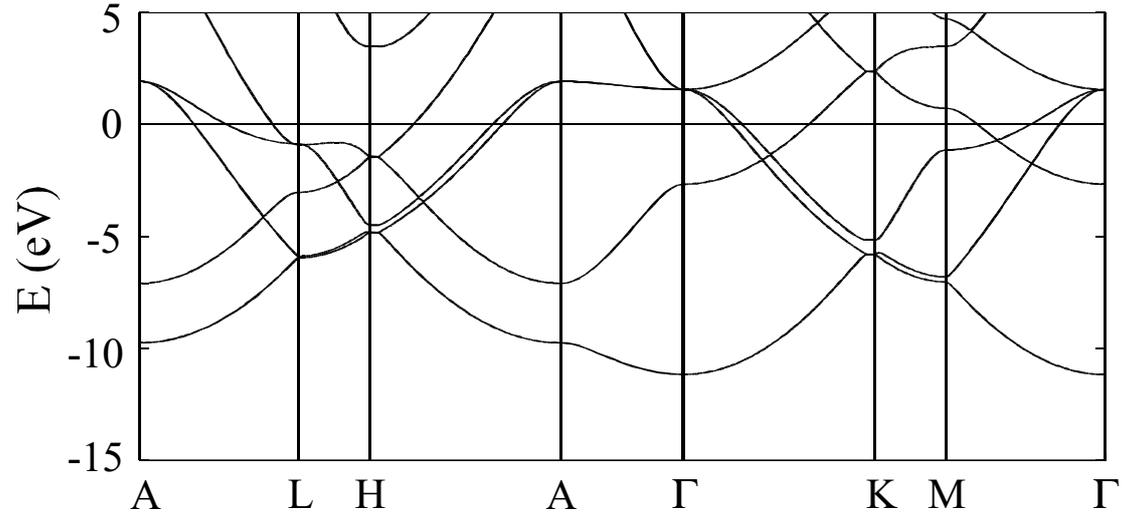
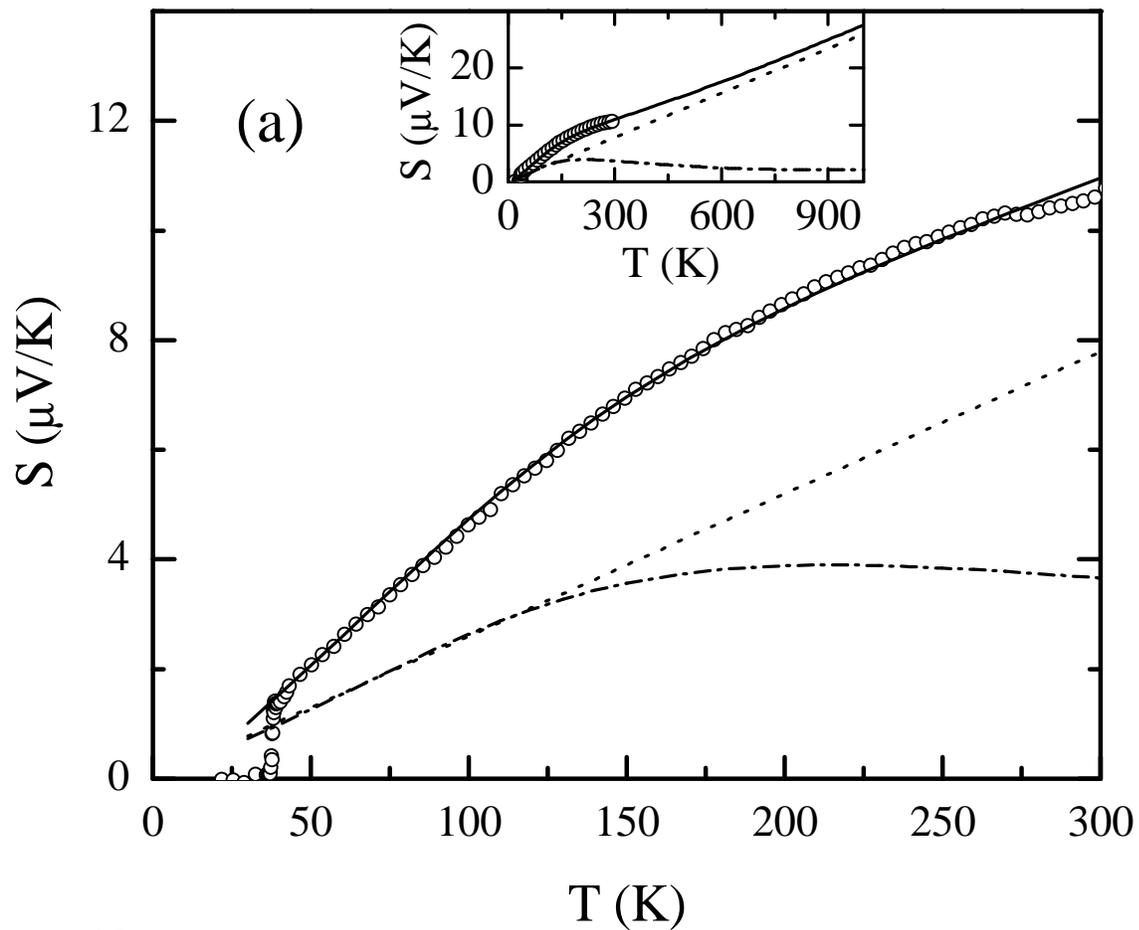
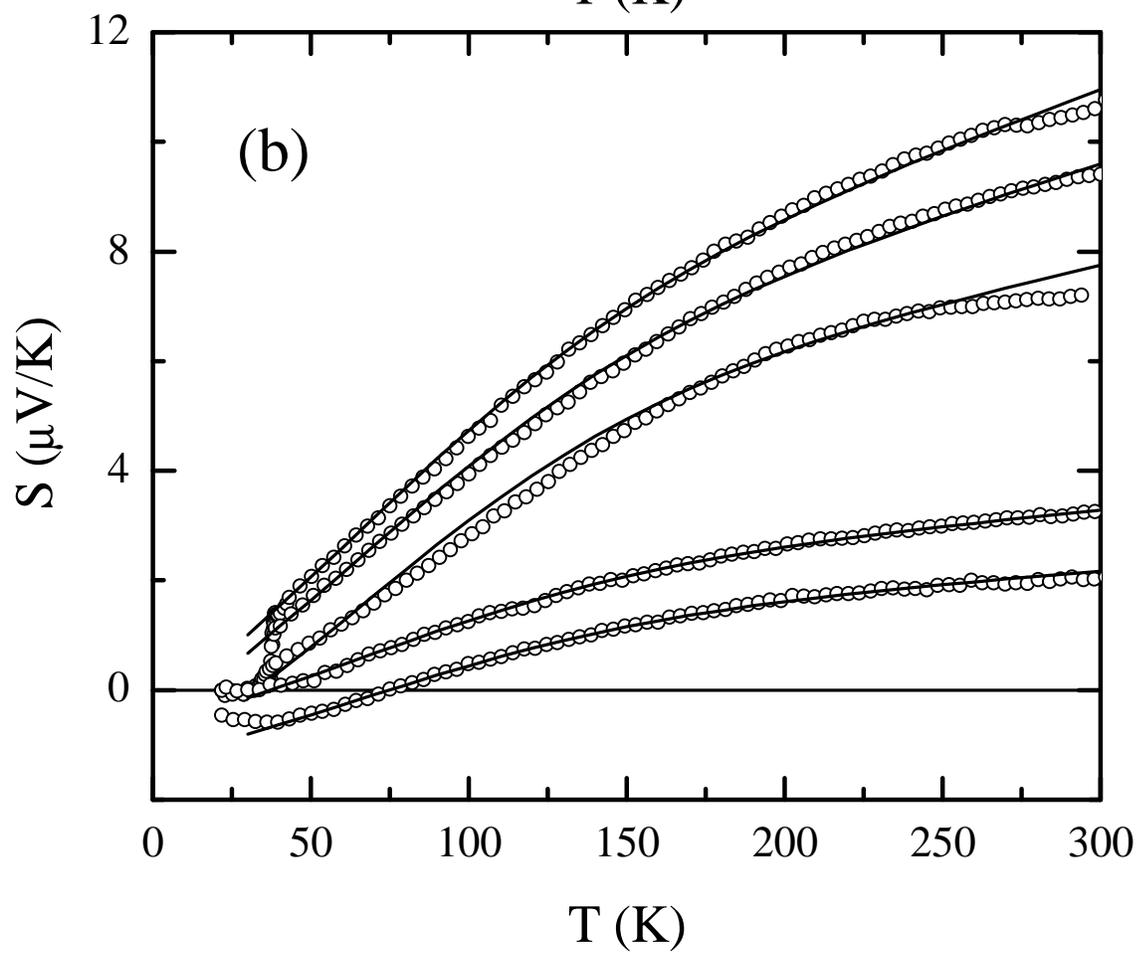
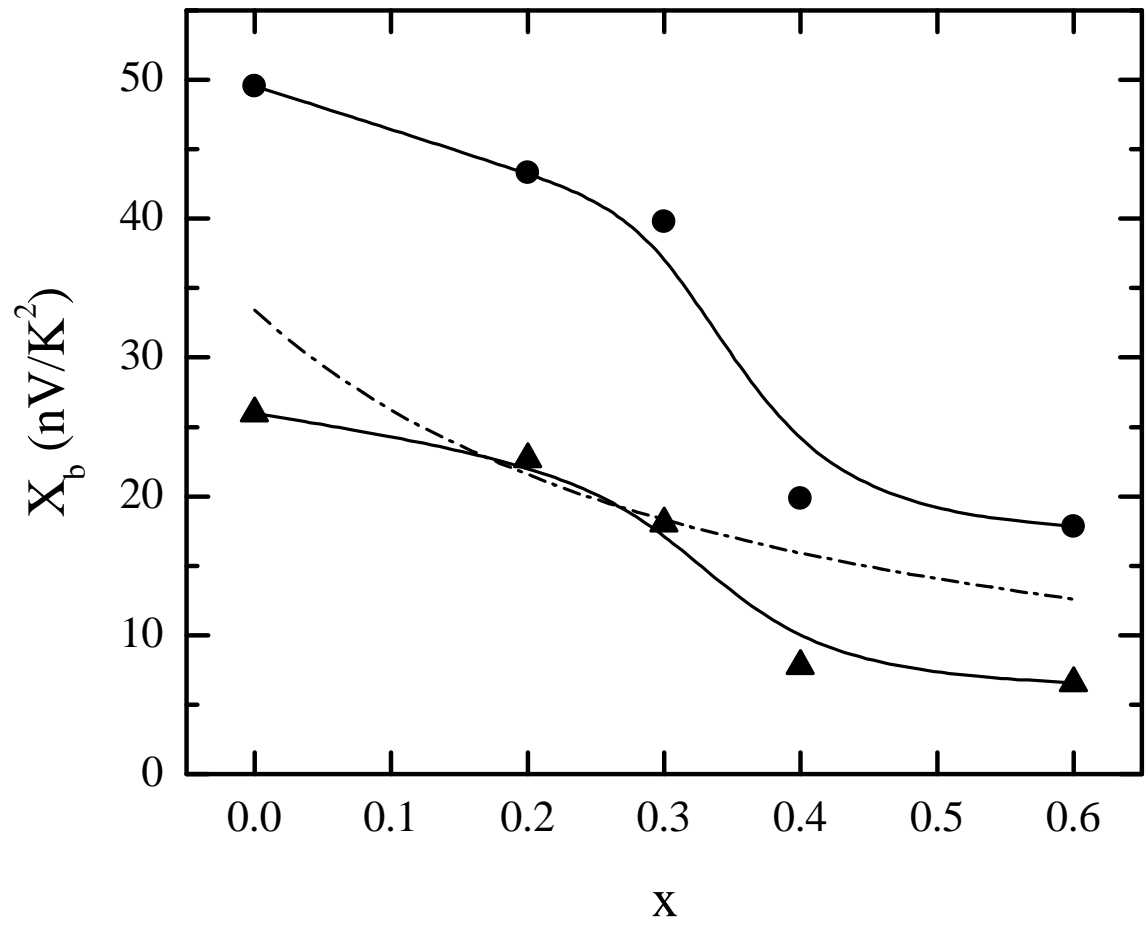
\end{document}